\documentclass[twocolumn,showpacs,superscriptaddress]{revtex4-1}
\usepackage{multirow,amssymb,amsbsy,amsmath}
\usepackage{graphicx}
\usepackage{verbatim}
\usepackage[colorlinks=true,linkcolor=black]{hyperref}
 
\def\ket#1{\mathinner{|{#1}\rangle}}

\begin{document}

\title{Correlations in quantum states and the local creation of quantum discord}

\author{Manuel Gessner}
\email{manuel.gessner@physik.uni-freiburg.de}
\affiliation{Physikalisches Institut, Universit\"at Freiburg, Hermann-Herder-Strasse 3, D-79104 Freiburg, Germany}

\author{Elsi-Mari Laine}
\email{emelai@utu.fi}
\affiliation{Turku Centre for Quantum Physics, Department of Physics and Astronomy, University of Turku, FI-20014 Turun yliopisto, Finland}

\author{Heinz-Peter Breuer}
\affiliation{Physikalisches Institut, Universit\"at Freiburg, Hermann-Herder-Strasse 3, D-79104 Freiburg, Germany}

\author{Jyrki Piilo}
\affiliation{Turku Centre for Quantum Physics, Department of Physics and Astronomy, University of Turku, FI-20014 Turun yliopisto, Finland}

\date{\today}

\begin{abstract}
Quantum discord is usually referred to as a measure for quantum correlations. In the search of the fundamental resource to gain a quantum advantage in quantum information applications, quantum discord is considered a promising candidate. In this paper we present an alternative view on quantum correlations in terms of the rank of the correlation matrix as introduced in [Phys. Rev. Lett. 105, 190502 (2010)]. According to our analysis, information about the quantum discord does not necessarily determine the amount of quantum correlations but rather the quantumness of the state. Nonzero quantum discord is only a necessary but not a sufficient condition for correlations above the classically achievable limit. This becomes clear when we consider states of nonzero discord which can be created from zero discord states only by a single local operation. We further show that the set of these states has measure zero.
\end{abstract}

\pacs{03.65.Ud, 03.65.Yz, 03.67.Mn}

\maketitle

The efficient usage of quantum correlations has led to some of the most striking discoveries of modern physics \cite{HHHH}. However, a clear identification of the basic resource for such applications has not yet been achieved. Popular concepts, such as teleportation of quantum states \cite{Bennett93}, quantum cryptography \cite{Ekert,Bennett}, quantum dense coding \cite{Bennett92} and effective algorithms for quantum computers \cite{Feynman,Shor} have been developed employing quantum entanglement. However, some computational tasks are expected to gain advantage with respect to classical limits even without the usage of entanglement. These tasks get their advantage from quantum discord, which is generally regarded as a measure for quantum correlations \cite{KnillLaflamme,Datta,Henderson,Zurek,Modi}. In general, the fundamental resource for the quantum advantage as well as the interpretation of quantum discord are still under debate.

In a recent paper Daki\'c, Vedral and Brukner point out that a local operation can create a state of nonzero quantum discord from a classical state \cite{DVB}, which has zero discord by definition: The local operation $|0\rangle\rightarrow|\varphi_0\rangle$, $|1\rangle\rightarrow|\varphi_1\rangle$ applied to the first subsystem of the classical state $1/2(|00\rangle\langle 00|+|11\rangle\langle 11|)$ produces the state $1/2(|\varphi_00\rangle\langle \varphi_00|+|\varphi_11\rangle\langle \varphi_11|)$, which has nonzero discord as long as $\langle\varphi_0|\varphi_1\rangle\neq0$ \cite{DVB}. Since local operations do not increase the amount of correlations between two subsystems, we expect the correlations of the resulting state to be lower than or equal to the correlations of the classical state -- a statement which is not reflected by the quantum discord. In general, the local creation of quantum discord has drawn a lot of interest recently \cite{Ciccarello,Hu-a,Bruss,Hu-b,Paternostro}.

In the same paper \cite{DVB}, the authors introduce the rank of the correlation matrix, henceforth denoted as $L$, which is a uniquely defined discrete number indicating the minimal amount of bipartite product operators needed in a linear combination to represent a given quantum state. Only product states have $L=1$ and for a higher rank the state is correlated. For classical states, an upper bound for $L$ can be derived and in order to achieve values above this bound, genuine quantum correlations are required. Further, we will show in this paper, that $L$ can not increase under local operations. Since local operations can change the properties of the subsystems in a way that a classical state is converted into a nonzero discord state, but can never increase the correlations in terms of $L$, we take the view that besides the quantum discord also $L$ is needed in order to characterize quantum correlations. A natural question to ask here is, how many of the nonzero discord states can be reached from classical states by applying only local operations? In this paper we explore this border and show that only a set of measure zero can be created locally from classical states.

Consider a bipartite Hilbert space $\mathcal{H}_A\otimes\mathcal{H}_B$ formed by the two subspaces $\mathcal{H}_A$ and $\mathcal{H}_B$ with dimensions $d_A$ and $d_B$, respectively. A classically correlated state is of the form
\begin{align}
 \rho_{c}=\sum_{i=1}^{d_A}\sum_{j=1}^{d_B}p_{ij}\Pi_i^A\otimes\Pi_j^B,
\label{clcl}
\end{align}
with a joint probability distribution $\{p_{ij}\}$ and the complete sets of rank-one projectors $\Pi_i^A=|i\rangle\langle i|$, $\Pi_j^B=|j\rangle\langle j|$, where $\{|i\rangle\}$ and $\{|j\rangle\}$ form an orthogonal basis of $\mathcal{H}_A$ and $\mathcal{H}_B$, respectively. Clearly, this state is just an operator formalism of the classical probability distribution $p_{ij}$ without any quantum nature. 

For the definition of the correlation matrix we assume that a given state $\rho$ is represented in terms of arbitrary bases of Hermitian operators $\{A_i\}$ and $\{B_i\}$ of the two Hilbert spaces $\mathcal{H}_A$ and $\mathcal{H}_B$, respectively:
\begin{align}
 \rho=\sum_{i=1}^{d_A^2}\sum_{j=1}^{d_B^2}r_{ij}A_i\otimes B_j.
\end{align}
The real-valued $d_A^2\times d_B^2$ correlation matrix $R=(r_{ij})$ can be decomposed into its singular value decomposition with the aid of orthogonal matrices $U$ and $V$ as $R=U\text{diag}(c_1,c_2,\dots,c_L,0,\dots)V^T$, where the $c_i$ are the nonzero singular values and $L$ denotes the rank of the correlation matrix of $\rho$. Introducing the matrices $S_i=\sum_ju_{ji}A_j$ and $F_i=\sum_jv_{ji}B_j$, where $U=(u_{ij})$ and $V=(v_{ij})$, the state $\rho$ can be written in a diagonal representation as \cite{DVB}
\begin{align}
 \rho=\sum_{i=1}^Lc_iS_i\otimes F_i.
\label{diagonalstate}
\end{align}
The state $\rho$ is then characterized by two different properties. On the one hand, $L$ determines how many product operators are needed to represent $\rho$ and thereby allows to draw conclusions about the total correlations. On the other hand, the commutation relations of the local operators $S_i$ determine the local quantumness of the state in the subsystem $\mathcal{H}_A$, indicated by the quantum discord \cite{DVB}: A necessary and sufficient condition for zero quantum discord is given by $[S_i,S_j]=0$ for all $i,j=1,\dots,L$ \cite{DVB}. Analogously, the commutation relation of the $F_i$ determine the quantum discord with respect to subsystem $\mathcal{H}_B$. Based on this insight a discord witness which only requires knowledge of $L$ was proposed in Ref.~\cite{DVB}. For classical states, the upper limit for $L$ is given by the minimum of the subsystem dimensions  $d_{\text{min}}=\text{min}\{d_A,d_B\}$, whereas general quantum states can achieve all values of $L\leq d_{\text{min}}^2$. Hence, a state of nonzero discord can be identified by $L>d_{\text{min}}$ \cite{DVB}. However, for states with nonzero discord which can be created locally from a classical state we have $L\leq d_\text{min}$, as we will show later. Thus the locally created nonzero discord states form a special class within all nonzero discord states: the rank of their correlation matrix $L$ is compatible with a classical state. We will now consider an example which illustrates, why the discord alone cannot provide a good measure for quantum correlations, but also the knowledge about $L$ is needed.

We investigate two states of two qubits. First we consider a Werner state,
\begin{align}
 \rho_W=\frac{1-z}{4} I+z|\Psi\rangle\langle\Psi|,
\end{align}
with the Bell state $|\Psi\rangle=1/\sqrt{2}(|00\rangle+|11\rangle)$. We choose $z=1/3$, making it a separable state. The rank of the correlation matrix for this state is maximal, i.e., $L=4$. Using the results of Ref.~\cite{Luo} the exact quantum discord according to the original definition by Ollivier and Zurek \cite{Zurek} can be obtained analytically for this state. We obtain $\mathcal{D}(\rho_W)=0.125815$. Now consider the state
\begin{align}
 \rho_L=\frac{1}{2}(|00\rangle\langle00|+|\!+1\rangle\langle+1|),
\end{align}
with $\ket{+}=1/\sqrt{2}(\ket{0}+\ket{1})$.The state can be created locally from the classical state
\begin{align}
 \rho_C=\frac{1}{2}(|00\rangle\langle00|+|11\rangle\langle11|)
\end{align}
by the operation $\Phi\otimes I$, with $\Phi(X)={|0\rangle\langle0|X|0\rangle\langle0|}+{|+\rangle\langle1|X|1\rangle\langle+|}$. Clearly, the state has $L=2$, compatible with a classical state. The quantum discord of this state can be obtained analytically using the results of Ref.~\cite{Shi}. We find that $\mathcal{D}(\rho_L)=0.201752$ which is higher than the value of the separable Werner state. For a reasonable measure of quantum correlations, it would be expected that a state which can be locally created from a classically correlated state would be less quantum correlated than a state which cannot be created with local operations only. This is not reflected by the quantum discord alone which therefore cannot be used as a single quantifier of quantum correlations. However, local operations cannot increase the rank of the correlation matrix. If a local operation is applied to a state of the form (\ref{diagonalstate}) one obtains $(\Phi\otimes I)\rho=\sum_{i=1}^Lc_i\Phi(S_i)\otimes F_i$. We represent $\Phi(S_i)$ in a linear combination of the original basis of matrices $\Phi(S_i)=\sum_{k=1}^{d_A^2}d_{ik}A_k$. The rank of the correlation matrix after application of this local operation is given by the rank of $(c_id_{ik})$. This in turn is given by the minimum of $L$ and the rank of $D=(d_{ik})$, which cannot be larger than $L$. As a consequence none of the nonzero discord states with $L>d_\text{min}$ can be created from classical states with local operations: $L$ detects nonzero discord states with high correlations which are clearly beyond the reach of classical states. If a quantum advantage cannot be achieved with classical states, the entire class of locally producible nonzero discord states can be excluded from the set of possibly useful states in computational tasks. Otherwise one could just take the local operation as part of the protocol and start out with a classical state. This might help to understand the important question why discord leads to a computational advantage in some situations but does not in others \cite{Modi}.

We will now identify the set of locally producible states, and show that this is indeed a very special class of states within the nonzero discord states: it has measure zero. By applying a local operation of the form $\Phi_A\otimes \Phi_B$ to a classical state $\rho_c$ of the form of Eq.~(\ref{clcl}) it is possible to generate any arbitrary state of the form
\begin{align}
 \rho=\sum\limits_{i=1}^{d_\text{min}}p_i\rho^A_i\otimes\rho_i^B
\label{locallyproducible}
\end{align}
To see this, let us first consider a subset of classical states defined in  Eq.~(\ref{clcl}) for which $p_{ij}=\delta_{ij}\tilde{p}_{i}$. The state can be written in the form
\begin{align}
 \tilde{\rho}_{c}=\sum_{i=1}^{d_\text{min}}\tilde{p}_{i}\Pi_i^A\otimes\Pi_i^B,
\label{clcls}
\end{align}
Now let us assume that the state $\rho_i^A$ has a spectral decomposition $\rho_i^A=\sum_{j=1}^{d_A}w^i_j|\Psi^i_j\rangle\langle\Psi^i_j|$. We define the Kraus operators $K_{ij}=\sqrt{w^i_j}|\Psi^i_j\rangle\langle i|$ which lead to the completely positive and trace preserving map
\begin{align}
 \Phi(X)=\sum_{i,j=1}^{d_A}K_{ij}XK_{ij}^{\dagger}.
 \label{map}
\end{align}
Local application of this map to the rank-1 operator $\Pi_i^A$ yields the state $\rho_i^A$. Thus operations of the form given in Eq.~(\ref{map}) will yield an arbitrary state from a projection operator. By defining the map $\Phi_B$ in the same way we obtain
\begin{align}
 (\Phi_A\otimes \Phi_B)\tilde{\rho}_c=\sum_{i=1}^{d_\text{min}}\tilde{p}_{i}\rho_i^A\otimes\rho_i^B,
 \end{align}
Thus from the states of the form given by Eq.~(\ref{clcls}) we can create all states of the form of Eq.~(\ref{locallyproducible}) by local operations. Even from a general classical state given by Eq.~(\ref{clcl}) one can produce only states of this form since the classical state can be cast into the form 
\begin{align}
\rho_{c}=\sum_{i=1}^{d_\text{min}}p_{i}\Pi_i^A\otimes\hat{\rho}_i^B,
\label{clcl2}
\end{align}
with $p_i\hat{\rho}_i^B=\sum_jp_{ij}\Pi_j^B$. A generalization to trace-decreasing maps, as e.g. a selective measurement, does not generate further possibilities for the outcome. Thus we find that by local operations we cannot create a general separable state from a classical one. However, all states which can be presented as a convex sum of $d_{\text{min}}$ product states can be created locally.

Next we show that the set of states of the form (\ref{locallyproducible}) is of measure zero. We begin with the class of separable states which can be written as a convex linear combination of $s$ product states:
\begin{align}
 \rho=\sum_{i=1}^sp_i\rho_i^A\otimes\rho_i^B.
\label{sproductstates}
\end{align}
A general density matrix of a $d$-dimensional Hilbert space can be described by $d^2-1$ real parameters. The states (\ref{sproductstates}) consist of $s$ product states, requiring in total an amount of $s(d_A^2-1+d_B^2-1)$ real parameters. Additionally, the probability distribution $p_i$ must be specified, which can be done by $s-1$ real parameters and the normalization condition. Hence, in order to describe the class of states defined in Eq.~(\ref{sproductstates}) no more than $s(d_A^2-1+d_B^2-1)+s-1$ real parameters are needed. On the other hand, a general state of the composite Hilbert space requires $d_A^2d_B^2-1$ real parameters. This implies that the set of states which can be described as a convex combination of no more than $s$ product states has Lebesgue-measure zero as long as it can be described by strictly less real parameters, yielding the condition
\begin{align}
 s(d_A^2+d_B^2-1)-1<d_A^2d_B^2-1.
\label{measurezeroineq}
\end{align}
The set of states which can be produced from truly classical states with local operations is given by the special case $s=\text{min}\{d_A,d_B\}$. Since the expression is symmetric in the variables $d_A$ and $d_B$ we can assume $d_A\leq d_B$ and check the validity of the inequality for $s=d_A$. The function
\begin{align}
 f(d_A,d_B)=d_A^2d_B^2-d_A^3-d_Ad_B^2+d_A
\end{align}
must be larger than zero for all values of $d_A\geq2$, $d_B\geq2$. By looking at the derivatives
\begin{align}
 \frac{\partial f(d_A,d_B)}{\partial d_A}&=(2d_A-1)d_B^2-3d_A^2+1\notag\\&\geq3(d_B^2-d_A^2)+1\geq0,\notag\\
 \frac{\partial f(d_A,d_B)}{\partial d_B}&=2d_B(d_A^2-d_A)>0,
\end{align}
we find that by increasing the values of $d_A$ and $d_B$, we can only increase the value of $f(d_A,d_B)$. Therefore, it suffices to consider the lowest possible values of $d_A=2$ and $d_B=2$. We have $f(2,2)=2>0$ showing that the inequality (\ref{measurezeroineq}) is fulfilled for $s=\text{min}\{d_A,d_B\}$ and the set of states which can be produced locally from classical states has measure zero. 

So far we have seen that it is possible to create any separable state which can be written as a convex sum of $d_{\text{min}}$ product states from a classical state with local operations. The minimal number $s_{\text{min}}$ of product states required to compose a given separable state is very hard to determine. On the other hand the rank $L$ of the correlation matrix is easy to obtain, even experimentally \cite{DVB}. How $L$ is connected to the number $s_{\text{min}}$ exemplifies the connection between locally created discord and compatibility of $L$ with classical states. The first question is whether all states with $L\leq d_{\text{min}}$, which are compatible with classical correlations, can be created locally from classical states. This it not the case, first of all since there exist entangled states with sufficiently low rank. The state $|\Psi\rangle=1/\sqrt{2}(|\varphi_0\varphi_0\rangle+|\varphi_1\varphi_1\rangle)$, where $\{|\varphi_i\rangle\}_{i=0}^4$ is an orthonormal basis of $\mathbb{C}^5$, has a rank $4$ correlation matrix, whereas the classical upper limit for the rank of states in the $\mathbb{C}^5\otimes\mathbb{C}^5$ Hilbert space is given by $5$. But are there any separable states with a low-rank correlation matrix which cannot be created locally from classical states? This question immediately leads to the problem of determining the minimal number of product states which are needed in the convex linear combination to represent a given separable state. Apart from special cases this question remains unanswered -- only upper and lower bounds could be derived \cite{HHHH,Sanpera,diVincenzo}. 

\begin{figure}
 \includegraphics[width=0.45\textwidth]{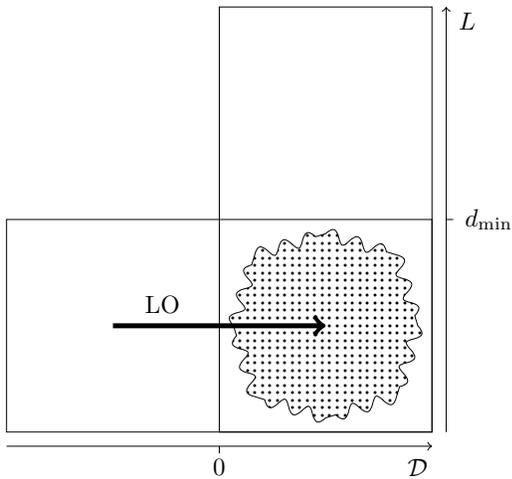}
 \caption{Schematic picture of the set of quantum states, classified by their quantum correlations which are characterized by the two quantities $\mathcal{D}$ (discord) and $L$. The correlations of the state are identified by $L$ whereas the quantum nature is indicated by $\mathcal{D}$. In order to have correlations above the classical upper limit of $d_{\text{min}}$ a nonzero value of $\mathcal{D}$ is required. The square on the bottom left represents the set of classical (zero discord) states. All of them have $L\leq d_{\text{min}}$. The top right square depicts the set of quantum correlated states in terms of $L$, since $L>d_{\text{min}}$. The bottom right square contains all states of nonzero discord but with low correlations, $L\leq d_{\text{min}}$. A subset of these, indicated by a dotted area, can be reached by local operations (LO) from the set of classical states.}
 \label{figstatespace}
\end{figure}

However, a connection between $L$ and the number $s_\text{min}$ of product states in the convex sum of a separable state can be derived. Let $\rho=\sum_{i=1}^sp_i\rho_i^A\otimes\rho_i^B$ be a state with correlation matrix rank $L$, where all of the $p_i$ are assumed to be nonzero. We prove that $L=s$ iff the states $\rho_i^A$ and $\rho_i^B$ are linearly independent. First of all, note that for all separable states $L\leq s$. If all the $\rho_i^A$ and $\rho_i^B$ are linearly independent they can be extended to a complete basis of Hermitian matrices in their respective spaces. The state then is already given in a diagonal representation and the rank $L$ of the correlation matrix is given by $s$. Next we look at the converse statement. Assume that a linear dependence of the form $\rho_s^A=\sum_{i=1}^{s-1}x_i\rho_i^A$ exists such that the states in the set $\{\rho_i^A\}_{i=1}^{s-1}$, as well as in $\{\rho_i^B\}$ are linearly independent. We can rewrite the state as
\begin{align}
 \rho&=\sum_{i=1}^{s-1}\left[p_i\rho_i^A\otimes\rho_i^B+p_sx_i\rho^A_i\otimes\rho^B_s\right]\notag\\
     &=\sum_{i=1}^{s-1}\rho_i^A\otimes(p_i\rho_i^B+p_sx_i\rho_s^B).
\label{reducedset}
\end{align}
The linear independent operators $X_i^B=p_i\rho_i^B+p_sx_i\rho_s^B$ can be extended to a complete basis of $\mathcal{H}_B$. The tensor product of these with the extension of the linear independent set $\{\rho_i^A\}_{i=1}^{s-1}$ forms a basis of matrices on $\mathcal{H}_A\otimes\mathcal{H}_B$ and from Eq.~(\ref{reducedset}) it can be immediately seen that $L=s-1$. This can easily be extended to the case of linear dependencies in the set of $\rho_i^B$ or linear dependencies of more than one element. However, the $X_i^B$ are not in general quantum states. A sufficient but not necessary condition for the possibility to express the state $\rho$ as a linear combination of less than $s$ quantum states is that all the $x_k$ are positive. Thus, the set of states that can be obtained from classical states with local operations is a subset of the set of separable states with $L\leq d_\text{min}$. The structure of the state space in terms of the rank and quantum discord is illustrated in Fig. \ref{figstatespace}.

Our results suggest that quantum discord should be interpreted as the local quantumness of the state, but does not necessarily make a statement about the amount of quantum correlations. This is reflected by the fact that quantum discord increases under local operations. Since the rank of the correlation matrix $L$ cannot increase under local operations we suggest that an appropriate measure for quantum correlations should measure also $L$. This was demonstrated by an example of two states, where the first state has a maximal $L$ whereas the other state has an $L$ compatible with a classical state. However, the state with maximal $L$ has a smaller value of the discord than the locally producible state.

We have also studied the properties of the set of states with $L$ compatible with classical states. We found that states, which are locally accessible from classical states form a subset of this set with measure zero. Thus by local operations it is possible to produce only a very restricted class of nonzero discord states. A question still remaining is whether all separable states which have $L$ compatible with a classical state can actually be created locally from a classical state.

\textit{Acknowledgements.} This work was supported by the German Academic Exchange Service (DAAD), the Graduate School of Modern Optics and Photonics, Academy of Finland (Project 259827), the Jenny and Antti Wihuri Foundation and the Magnus Ehrnrooth Foundation.

\end{document}